\documentclass[prl,twocolumn,superscriptaddress,aps,amsmath,amsfonts,notitlepage,longbibliography]{revtex4-2}
\usepackage{amsmath,amsfonts,amssymb,dsfont,graphicx,bm,caption,subcaption}
\graphicspath{{Figures/}}
\usepackage{color}
\usepackage{hyperref}
\usepackage{tikz}
\usepackage{pbox}

\def\be{\begin{equation}}
\def\ee{\end{equation}}
\def\bea{\begin{eqnarray}}
\def\eea{\end{eqnarray}}
\def\bpm{\begin{pmatrix}}
\def\epm{\end{pmatrix}}

\def\Im{\mathop{\rm Im}}

{\makeatletter
\def\ps@pprintTitle{
	\let\@oddhead\@empty
	\let\@evenhead\@empty
	\def\@oddfoot{\centerline{\thepage}}%
	\let\@evenfoot\@oddfoot}
\makeatother}

\begin{document}
\title{Non-Fermi-liquid behavior from cavity electromagnetic vacuum fluctuations at the superradiant transition}
\author{Peng Rao}
\affiliation{Max Planck Institute for the Physics of Complex Systems, N\"othnitzerstr. 38, 01187 Dresden, Germany}
\author{Francesco Piazza}
\affiliation{Max Planck Institute for the Physics of Complex Systems, N\"othnitzerstr. 38, 01187 Dresden, Germany}
\date{\today}
\begin{abstract}
We study two-dimensional materials where electrons are coupled to the vacuum electromagnetic field of a cavity. We show that, at the onset of the superradiant phase transition towards a macroscopic photon occupation of the cavity, the critical electromagnetic fluctuations, consisting of photons strongly overdamped by their interaction with electrons, can in turn lead to the absence of electronic quasiparticles. Since transverse photons couple to the electronic current, the appearance of non-Fermi-Liquid behavior strongly depends on the lattice. In particular, we find that in a square lattice the phase space for electron-photon scattering is reduced in such a way to preserve the quasiparticles, while in a honeycomb lattice the latter are removed due to a non-analytical frequency dependence of the damping $\propto |\omega|^{2/3}$. Standard cavity probes could allow to measure the characteristic frequency spectrum of the overdamped critical electromagnetic modes responsible for the non-Fermi-liquid behavior.
\end{abstract}

\maketitle

\emph{Introduction.---}
Certain strongly correlated metals do not behave according to Landau's Fermi-liquid theory (see \cite{chowhury_2018} for a recent classification). In most of the cases this happens in correspondence to a quantum critical point separating a normal metallic phase and a symmetry-broken phase \cite{loenheisen_2007,stewart_2001}. Within this scenario which has special relevance in two-dimensional materials, the strong coupling between the Fermi surface and critical order-parameter fluctuations leads to the loss of Landau's quasiparticles and thus to non-Fermi-Liquid behavior \cite{altshuler_1994,kim_1994,abanov2003quantum,chubukov2003nonanalytic,metzner_2003,sslee_2009,sslee_2013,mandal_2015,metlitski_2010a,metlitski_2010b,samokhin_2006,Holder2014,piazza_2016,pimenov_2018,sslee_2018,driskell2021identification}. A direct signature is a non-analytical frequency dependence of the quasiparticle damping $\sim |\omega|^\alpha$, with $\alpha<1$, as opposed to the usual Fermi-liquid damping $\sim \omega^2$ which becomes instead increasingly irrelevant towards the Fermi surface $\omega\to 0$. 
In order to improve our understanding of the emergence of non-Fermi-liquid behavior and its experimental relevance, it would be highly desirable to determine the microscopic origin of the bosonic degree of freedom whose critical fluctuations are responsible for removing the electronic quasiparticles.

The recent experimental possibility to strongly couple electrons in two-dimensional materials with photons confined within cavities \cite{garcia2021manipulating,schlawin2021cavity} has opened new avenues for controlling collective electronic phenomena and explore them in novel scenarios. These include cavity induced superconductivity \cite{Colella2018Quantum,Sentef2018,Schlawin2019,Schlawin2019a,Curtis2019,Thomas2019,sheikan_2019,li_manipulating_2020,Chakraborty_2021}, magnetism \cite{Mivehvar2019Cavity,Kiffner2019,Sentef2020,chiocchetta_2021,curtis_2022}, ferroelectricity \cite{Ashida2020,latini_2021}, as well as topological phenomena \cite{Pan2015Topological,dora_2015,Zheng2016Superradiance,Kollath2016Ultracold,Mivehvar2017Superradiant,sentef_hall_2019,appugliese2022breakdown,rokaj_2022}, that could be realized by engineering the quantum electrodynamical (QED) coupling between matter and light. These scenarios can also be investigated with synthetic matter made of ultracold atoms \cite{mivehvar_2021}.

In this work, we show that cavity QED within two-dimensional materials offers ideal conditions to implement and observe non-Fermi-liquid behavior. The fluctuations of the emergent bosonic degree of freedom which induce non-Fermi-liquid behavior in the standard scenario are here substituted by the fluctuations of the vacuum electromagnetic field, i.e., a microscopic degree of freedom whose dynamics and coupling with electrons can be controlled by cavity engineering. 
Moreover, two-dimensional materials within layered structures \cite{schlawin2021cavity} (and even more so synthetic ultracold-atomic systems \cite{mivehvar_2021,Roux2020,Zhang2021}) offer an enhanced tuneability of electronic properties, including Coulomb interactions as well as the role of impurities and phonons. This potentially allows to realize a situation where the QED coupling with cavity photons is dominant. Cavity mirrors create a gap in the electromagnetic spectrum. However, electromagnetic modes can be made critical by reaching the transition point towards superradiance \cite{Piazza2014Umklapp,Keeling2014Fermionic,Chen2014Superradiance,nataf_2019,andolina_2019,rabl_2020,Guerci2020,Andolina2020}, at which the hybridization with the matter creates gapless polariton modes.
Since the QED coupling depends on the electron momentum, we find that non-Fermi-liquid behavior can be controlled via the choice of the underlying lattice. We consider here a square and a honeycomb lattice away from unit filling. While in both cases the hybridization with matter leads to overdamped polaritons at the critical point, the phase space for electron-photon scattering is such that non-Fermi-liquid behaviour is absent for the square lattice, where the electron quasiparticle damping is $\propto \omega^2 \log|\omega|$, but present for the honeycomb lattice, where the damping is $\propto |\omega|^{2/3}$. Measurements of the cavity spectrum could show, instead of a well-defined resonance, non-analytical power-law tails revealing the presence of the critical bosonic fluctuations responsible for the non-Fermi-liquid behavior. The present scenario can also be realized with ultracold fermionic atoms in confocal cavities \cite{rylands2020photon}.


\begin{figure}[t]
\centering
\includegraphics[width=\linewidth]{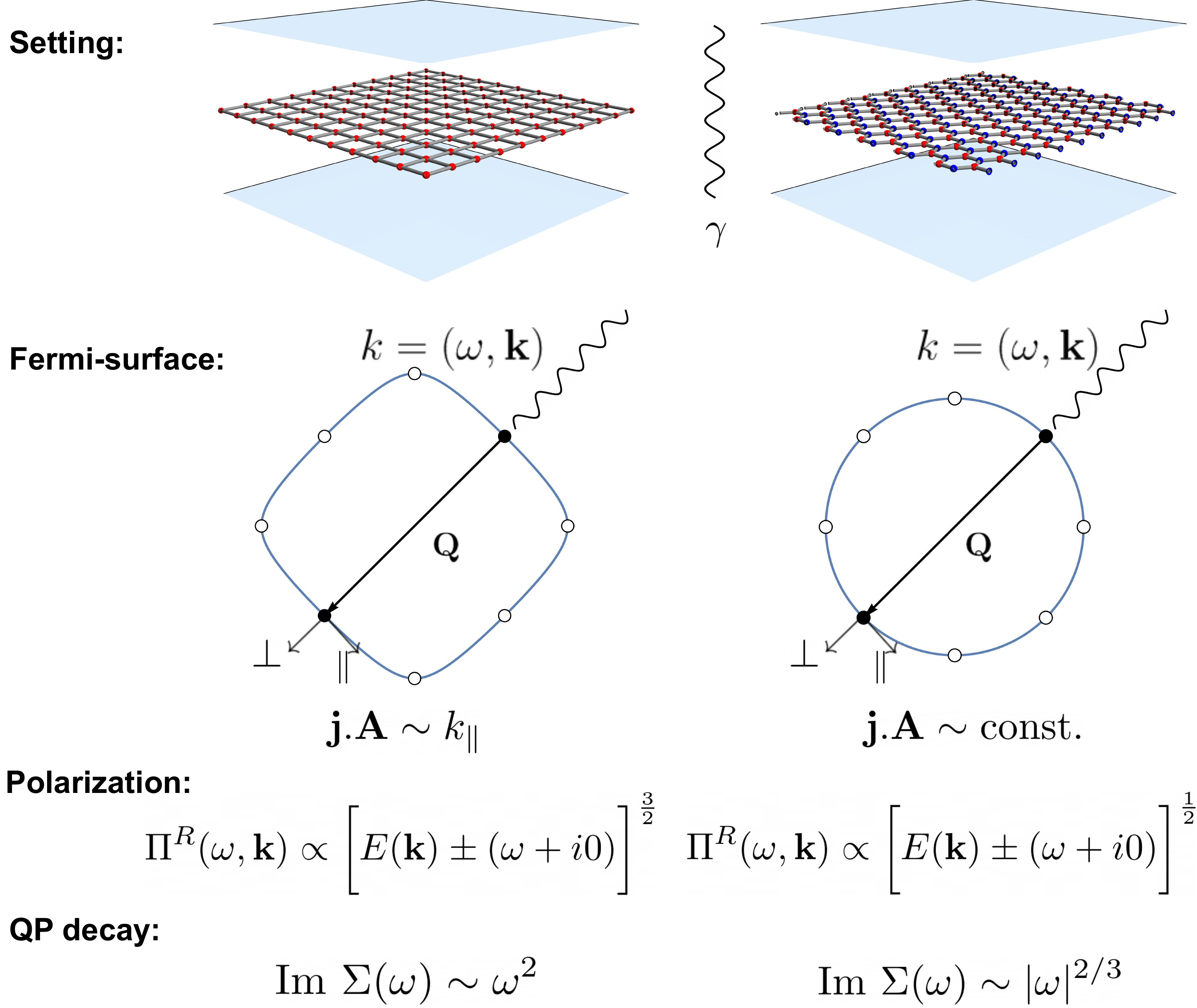}
\caption{
Summary of the main results for the two cases considered: cavity photons coupled to electrons on a two-dimensional square lattice (left column) and honeycomb lattice (right column, red and blue atoms belong to the $A$, $B$ sublattices). Potential hot-spots are shown in white circles. For each lattice we show explicitly nesting momenta $\mathbf{Q}$ for one pair of hot-spots in black dots. Note that on the honeycomb lattice, the entire fermi-surface is `hot'.
}
\label{fig:lattice-geometry}
\end{figure}

\emph{Model.---}
We consider the cavity consisting of two parallel perfect conducting mirrors and choose the Coulomb gauge: $\text{div} \ \mathbf{A}=0$ for the electromagnetic field. The two-dimensional electronic system is placed in the middle of the two plates as illustraded in Fig.~\ref{fig:lattice-geometry}. Due to the boundary condition at the mirrors, the $n$-th photon modes acquires a finite `mass' $\omega_{0n}=n\pi/L$ where $L$ is the distance between two plates~\cite{Kakazu1994}. For simplicity we shall consider only the $n=1$ modes. Since, as we shall see, the relevant photon momenta for non-Fermi-liquid scaling are of the order of the inverse lattice constant $1/a$, this approximation is justified only for ultra-cold gases; for solid state systems $1/a$ instead exceeds the photon mass scale. However, as shown in the Supplemental Material~\cite{SM}, the inclusion of all cavity modes does not change the non-Fermi-liquid scaling. The photon dispersion is:
\begin{equation}\label{eq:PhotonDispersion}
    \omega_k^2 = k^2 + \omega_0^2,
\end{equation}
where $\mathbf{k}$ is the 2D photon momentum in the plane of the lattice. The free particle Hamiltonian is:
\begin{equation}\label{eq:ModelHamiltonian}
H_0 = - t \sum_{i,j,\alpha = \pm1} c^\dagger_{\alpha,i} c_{\alpha,j} + \sum_k \omega_k \left( b^\dagger_k b_k+\frac{1}{2}\right).
\end{equation}
Here $i, j$ are lattice sites and $\alpha$ is the spin index. The first term in (\ref{eq:ModelHamiltonian}) is the tight-binding electron Hamiltonian, and the second term describes the transverse photon modes. The free transverse photon Green's function has the form~\cite{Kakazu1994,SM}:
\begin{equation}\label{eq:PhotonGreensFunction}
    D_{0,ij}(\omega,\mathbf{k}) = -\frac{2}{L}\frac{1}{\omega^2-\omega_k^2}e_i(\mathbf{k}) e^{*}_j(\mathbf{k}),
\end{equation}
where $\mathbf{e}(\mathbf{k})$ is the transverse polarization vector: $\mathbf{k}.\mathbf{e}(\mathbf{k})=0$. 
Electron-photon coupling is introduced via the Peierls substitution: $c_i\rightarrow c_i \exp \int^{\mathbf{r}_i}_{\mathbf{r}_0} \mathbf{A}.d\mathbf{r}$ in $H_0$~\footnote{
Under the Coulomb gauge, the scalar potential $\phi$ mediates direct Coulomb interactions between electrons. Since Coulomb interactions are not expected to generate non-FL behavior, we shall neglect $\phi$ in this paper.}. For our purpose it is sufficient to keep the paramagnetic part linear in $\mathbf{A}$ \cite{Guerci2020}:
\begin{equation}\label{eq:ElectronPhotonCoupling}
V= - \sum_{\mathbf{k}}\mathbf{j}(\mathbf{k}).\mathbf{A}(\mathbf{k}), \ H = H_0 + V.
\end{equation}
In principle, higher-order terms are required for guaranteeing gauge invariance of the theory, but those do not influence the results in this work (see below).
At $\omega=0$, the longitudinal cavity mode decouples from electrons as a result of charge conservation $\mathbf{k}.\mathbf{j}=0$. Therefore, at small frequency, only the transverse cavity mode couples strongly to the Fermi-surface, whereas the longitudinal mode is suppressed by $\omega^2$. In what follows we shall only consider coupling between the transverse mode and electrons. 

\begin{figure}[t]
\centering
\includegraphics[width=\linewidth]{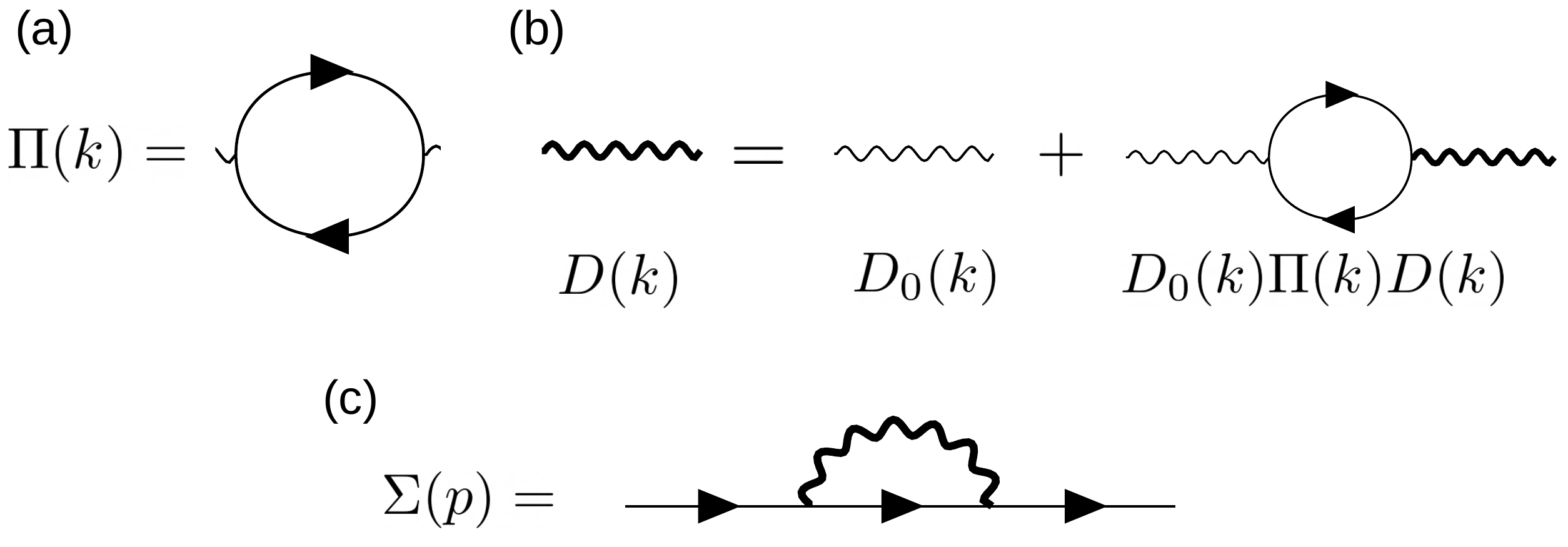}
\caption{
One loop diagrammatic representation for: (a) the Polarization function $\Pi(k)$; (b) the Dyson equation for photon propagator $D(k)$; and (c) electron self-energy $\Sigma(p)$ renormalized by the thick photon (polariton) line in (b).
}
\label{fig:feynman-diagrams}
\end{figure}

\emph{Superradiant critical point.---}
The light-matter coupling (\ref{eq:ElectronPhotonCoupling}) leads to hybridization between cavity photons and electronic excitations. The properties of the resulting polariton modes are described by the Dyson's equation for the retarded transverse photon Green's function (see Fig.~\ref{fig:feynman-diagrams}(b)): 
\begin{equation}\label{eq:DysonEquation}
    [D^R(k)]^{-1} = [D_{0}^R(k)]^{-1} - \Pi^R(k),
\end{equation}
with the polarisation function $\Pi^R(k)$. The superradiant critical point is reached at given momenta $\mathbf{k}$ such that the polariton gap closes: $\Pi^R(0,\mathbf{k})=[D_{0}^{R}(0,\mathbf{k})]^{-1}$. Beyond this point a macroscopic occupation of photons at momenta $\mathbf{k}$ becomes energetically favourable. For the case of multiple cavity modes, $D^R(k)$ becomes a matrix in cavity mode indices. The critical point is reached when one of the eigenvalues of $[D^R(0,\mathbf{k})]^{-1}$ first becomes zero~\cite{Andolina2020,SM}. Note that superradiance cannot happen at vanishing photon momenta which corresponds to a pure gauge. In this case the paramagnetic contribution is cancelled exactly by the diamagnetic part~\cite{Guerci2020,Andolina2020}.
In this work, we consider electrons on square and honeycomb lattices away from unit-filling, when the Fermi-surfaces have imperfect nesting. As we shall see next, superradiance is first reached at momenta $\mathbf{k}=\mathbf{Q}$, where $\mathbf{Q}$ is a nesting momentum.
We will show that at the critical point the polariton modes are strongly overdamped at small frequencies and near momenta $\mathbf{Q}$, and these can in turn destroy the electronic quasi-particles.

\emph{Square Lattice.---}
Let us start with the square lattice below unit filling. 
The current operator in the tight-binding approximation is given by:
\begin{equation}\label{eq:SquareLatticeCurrent}
\begin{split}
&\mathbf{j}(\mathbf{k})  =\sum_{\mathbf{p}}\mathbf{\Lambda}(\mathbf{p},\mathbf{k}) c^\dagger_{\mathbf{p}+\mathbf{k}}c_{\mathbf{p}},\\
\mathbf{\Lambda}(\mathbf{p},\mathbf{k}) &= -aet\sin\bigg[\bigg(\mathbf{p}+\frac{\mathbf{k}}{2}\bigg).\mathbf{a}_i\bigg] \frac{\mathbf{a}_i}{a}.
\end{split}
\end{equation}
Here $\mathbf{a}_i$ is the $i$-th lattice vector and $a$ is the lattice constant. There are special points on the Fermi surface which couple most strongly to the transverse photons. Generally, such a hot-spot point has a symmetric partner on the opposite side of the Fermi surface. They are connected by a nesting momentum $\mathbf{Q}$ which there lies perpendicular to the Fermi surface. On the square lattice away from unit filling, they are four pairs 
indicated in Fig.~\ref{fig:lattice-geometry}. The nesting momentum $\mathbf{Q}$ is only shown for one such pair in the figure.
At the hot spots the electron dispersion has the form (\ref{eq:HotSpotDispersion}):
\begin{equation}\label{eq:HotSpotDispersion}
\xi_{\pm}(\mathbf{k}) = \pm v_Fk_\perp + \frac{k_\parallel^2}{2m},
\end{equation}
where $k_\perp$ and $k_\parallel$ are components of $\mathbf{k}$ parallel and orthogonal to $\mathbf{Q}$, and $v_F$, $m$ are parameters of the Fermi-surface. We have assumed that we are sufficiently away from unit filling that the Fermi velocity $v_F$, which vanishes at the corners of the square Fermi-surface at unit filling, can be regarded as constant here.
In all cases the current-photon vertex that connect pairs of hot-spots in Eq.~(\ref{eq:SquareLatticeCurrent}) has the same form:
\begin{equation}\label{eq:HotSpotVertex}
\begin{split}
\mathbf{\Lambda}.\mathbf{e} \approx -a^2et\bigg(p_\parallel+\frac{k_\parallel}{2}\bigg).
\end{split}
\end{equation}
Here $\mathbf{p}$ is the quasimomentum measured from one of the hot-spots. The momentum-dependent vertex in (\ref{eq:HotSpotVertex}) affects the one-loop polarization function for the transverse cavity photons.
Given the superradiant critical point condition, $\Pi^R(0,\mathbf{Q})=[D^{R}_0(0,\mathbf{Q})]^{-1}$ in (\ref{eq:DysonEquation}), the expansion $\Pi^R(\omega,\mathbf{Q}+\mathbf{k})- \Pi^R(0,\mathbf{Q})$ at small $\omega$ and $|\mathbf{k}|$ contains the following non-analytical contribution from near the Fermi-surface:
\begin{equation}\label{eq:SquareLatticePolarisation1}
 -\frac{a^4(et)^2m^{\frac{3}{2}}}{6\pi v_F} \bigg( \bigg[E(\mathbf{k})+\omega+i0 \bigg]^{\frac{3}{2}}+ \bigg[E(\mathbf{k})-\omega-i0 \bigg]^{\frac{3}{2}}\bigg),
\end{equation}
where $E(\mathbf{k}) = v_Fk_\perp+k_\parallel^2/(4m)$. This term induces an imaginary part at large negative $E(\mathbf{k})$, corresponding to strong Landau damping of the polariton mode at the superradiant critical point:
\[
[-|E(\mathbf{k})| \pm (\omega +i0)]^{\frac{3}{2}} \rightarrow  \pm i [|E(\mathbf{k})|  \mp (\omega +i0)]^{\frac{3}{2}}.
\]

Subsequent terms in the expansion of $\Pi^R(k)- \Pi^R(0,\mathbf{Q})$ receive contributions far from the Fermi-surface and can be written in powers of $E(\mathbf{k})$. They provide dynamics for the polaritons. The retarded photon Green's function thus becomes:
\begin{equation}\label{eq:SquareLatticePhoton}
\begin{split}
&D^{R}(\omega,\mathbf{Q}+\mathbf{k}) = \bigg([E(\mathbf{k})+\omega+i0 ]^{\frac{3}{2}}+ \\ &+[E(\mathbf{k})-\omega-i0]^{\frac{3}{2}}+ b  E(\mathbf{k})+ c[E(\mathbf{k})]^2\bigg)^{-1},
\end{split}
\end{equation}
where we have included up to quadratic powers of $E(\mathbf{k})$ in the expansion~\cite{Holder2014}. The constants $b$ and $c$ are given by contributions far from the Fermi-surface. Note that the free photon dispersion (\ref{eq:PhotonDispersion}) gives in principle also powers of $\xi_+(\mathbf{k})$ in the denominator of $D^R$. These have been however neglected since they do not affect the quasi-particle decay, as will be justified below.

We now calculate the electronic quasi-particle decay as a function of frequency, due to the interaction with the Landau-damped polariton mode. The decay rate is given by twice the imaginary part of the electron self-energy, shown in Fig.~\ref{fig:feynman-diagrams}(c):
\begin{equation}\label{eq:SquareLatticeSelfEnergy}
\begin{split}
    &\Im \Sigma^R \left(\omega,\frac{\mathbf{Q}}{2}\right) \propto\int k_\parallel^2   \Im D^R[\xi_+(\mathbf{k})-\omega,\mathbf{k}-\mathbf{Q}] \frac{d^2k}{(2\pi)^2}
    \end{split}
\end{equation}
Here $0<\xi_+(\mathbf{k})<\omega$. In Eq.~(\ref{eq:SquareLatticeSelfEnergy}), the photon Green's function at small frequency and near momenta $\mathbf{Q}$ is considered, which corresponds to the polariton mode. We evaluate this integral in the Supplemental Material~\cite{SM}. Here we limit ourselves to highlighting the most important feature, namely that at sufficiently small $\omega$, the dominant contribution to (\ref{eq:SquareLatticeSelfEnergy}) comes from the region in the phase space in which polariton Landau-damping is strong: $-E(\mathbf{k})\gg \xi_+(\mathbf{k}) \sim \omega>0$. This justifies neglecting $\xi_+(\mathbf{k})$. The final result is:
\begin{equation}\label{eq:SquareLatticeSelfEnergy2}
\Im \Sigma^R\left( \omega, \frac{\mathbf{Q}}{2}\right) \sim \omega^2\log |\omega|.
\end{equation}
The electron quasiparticles remain thus well-defined in this case. 
This can be attributed to the momentum-dependence of the QED coupling \eqref{eq:ElectronPhotonCoupling} inherited from the electronic current operator \eqref{eq:SquareLatticeCurrent}, which suppresses the effect of the Landau-damped polaritons near the Fermi-surface (compare with Eq.~(\ref{eq:GrapheneSelfEnergy1}) on the honeycomb lattice below).

\emph{Honeycomb Lattice.---}
We now turn to the honeycomb lattice. For simplicity we shall focus on the $K'$-point in quasimomentum space. The single-particle Hamiltonian $H_0$ then has the form:
\begin{equation}\label{eq:GrapheneDispersion}
 H_0(\mathbf{k}) =  v_F \bm{\sigma}.\mathbf{k}.
\end{equation}
Here $\bm{\sigma}$ are Pauli matrices in pseudo-spin space and $\mathbf{k}$ is a 2D quasimomentum measured from $K'$. Eq.~(\ref{eq:GrapheneDispersion}) gives two particle-hole symmetric bands with energies $\varepsilon_{\lambda}(k)$ and eigenstates $u_\lambda(\mathbf{k})$:
\begin{equation}\label{eq:GrapheneEigenstates}
\varepsilon_{\lambda}(k) = \lambda v_Fk, \ u_\lambda(\mathbf{k}) = \frac{1}{\sqrt{2}}\begin{pmatrix} 1 \\ \lambda \exp(i\varphi_{\mathbf{k}})\end{pmatrix},
\end{equation}
where $\lambda = \pm$ and $\varphi_{\mathbf{k}}$ is the polar angle of $\mathbf{k}$. At finite positive chemical potential $\mu$, the Fermi-surface is a circle on the electron band. Contrary to the square lattice case, the current operator has no momentum dependence:
\begin{equation}\label{eq:GrapheneCurrent}
 \mathbf{j} = \frac{\delta}{\delta \mathbf{A}}H_0(\mathbf{k}-e\mathbf{A}) =  - ev_F \bm{\sigma}.
\end{equation}


Moreover, the whole Fermi-surface is nested, i.e., $\mathbf{Q} = 2\mathbf{p}_F$. 
At low-enough energies, we will assume each of the infinitely many hot-spot pairs to contribute independently \cite{metlitski_2010a} and therefore consider only one such pair, with dispersion given by Eq.~(\ref{eq:HotSpotDispersion}) with $m = p_F/v_F$. It gives a non-analytical contribution to the polarization function quantified by the expansion around the critical point $\Pi^R(\omega,\mathbf{Q}+\mathbf{k}) - \Pi^R(0,\mathbf{Q})$, which reads:
\begin{equation}\label{eq:GraphenePolarisation1}
\begin{split}
  \frac{e^2v_F\sqrt{m}}{2\pi } \bigg(\bigg[E(\mathbf{k})  +\omega +i0 \bigg]^{\frac{1}{2}}+\bigg[E(\mathbf{k}) - \omega -i0 \bigg]^{\frac{1}{2}}\bigg),
\end{split}
\end{equation}
where we have neglected terms of order $k^2/p_F^2$. The details of computing (\ref{eq:GraphenePolarisation1}) are given in the Supplemental Materials~\cite{SM}. In what follows, we include in (\ref{eq:GraphenePolarisation1}) also a linear term in $E(\mathbf{k})$ for polartion dynamics. As is the case for the square lattice, powers of $\xi_+$ do not need to be included. Therefore, for large negative $E(\mathbf{k})$, we find strongly Landau damped-polaritons at the superradiant critical point, albeit with a different non-analytic power law.
 
The quasiparticle decay rate at the hot-spots due to interactions with overdamped polaritons at small $\omega$ now becomes \cite{SM}:
\begin{equation}\label{eq:GrapheneSelfEnergy1}
\Im \Sigma^R\left( \omega, \mathbf{p}_F\right) \sim |\omega|^{2/3},
\end{equation}
Therefore, the quasi-particle decay exhibits non-FL scaling contrary to the square lattice case, despite the presence of strong Landau damping for both lattices. This is because, on the honeycomb lattice, the dominant contribution comes from the region $-E(\mathbf{k}) \sim \Lambda_0 (\omega/\Lambda_0)^{2/3}\gg \omega$, $\Lambda_0$ being a UV cut-off, whereas on the square lattice this region is suppressed by the additional powers of momenta from the current operator (\ref{eq:SquareLatticeCurrent}).

We note that the non-analytical form of the quasiparticle damping in \eqref{eq:GrapheneSelfEnergy1} is the same as predicted for non-commensurate charge-density-wave \cite{altshuler_1994,Holder2014} and Fulde-Ferrel-Larkin-Ovchinnikov \cite{piazza_2016,pimenov_2018} quantum critical points. The same damping form is found for 2D fermionic systems with emergent $U(1)$ gauge fields as well, despite the important physical difference consisting in the gauge propagator being Landau-damped at small frequencies and momenta~\cite{altshuler_1994,mandal2020,sslee_2009}.
We conclude this part by mentioning that the result \eqref{eq:GrapheneSelfEnergy1} does not affect the polariton Landau damping \eqref{eq:GraphenePolarisation1} which is induced by electrons at large momenta. This is typically the case for one-loop Eliashberg-type theories \cite{polchinski1994low}.

\emph{Implementation.---}
The light-matter coupling (\ref{eq:ElectronPhotonCoupling}) is relevant for electrons in solid state, which couple to photons via the current \cite{schlawin2021cavity}. For neutral atomic gases instead, the dispersive coupling between photons and the atomic center-of-mass is independent of the momentum of the latter \cite{mivehvar_2021}. Therefore, non-Fermi-liquid behavior of Eq.~\eqref{eq:GrapheneSelfEnergy1} will be present regardless of the lattice, as long as unit filling is avoided. The same situation can be engineered also in solid state using two-photon transitions exploiting the diamagnetic coupling \cite{Gao2020} or auxiliary electronic bands \cite{chiocchetta_2021}. 

\emph{Accessibility.---} The accessibility of our non-Fermi-liquid regime relies upon reaching the superradiant transition. In the solid-state microcavity setup, the related challenges have been addressed in Refs.~\cite{Guerci2020,Andolina2020}. Here one must also consider higher cavity modes $n\ne 1$, but the non-FL scaling remains unaffected~\cite{SM}. On the other hand, for ultracold Fermi gases superradiance has been already observed \cite{zhang2021observation}, albeit only with a single isolated cavity mode. The required extension to a 2D continuum of modes is available in confocal cavities \cite{kollar2015adjustable,rylands2020photon}.

\emph{Observability.---}
With the purpose of probing the non-Fermi-liquid behavior, layered structures allow to precisely measure the electronic spectral function of an embedded 2D material  using techniques like momentum-and-energy-resolved tunneling spectroscopy (MERTS) \cite{jang2017full} and thus potentially to directly probe the non-analytic behavior of the quasiparticle damping. Such measurements are standardly available also for neutral atomic gases through radio-frequency spectroscopy~\cite{punk_2007}.

Since in our scenario the bosonic order parameter fluctuations affecting the electrons are in fact photons, cavity probes can provide direct access to their characteristic overdamped behavior responsible in turn for the electronic quasiparticle damping.
At sufficiently low frequencies  $|\omega|<-E(\mathbf{k})$ , the usual resonant peaks in the cavity spectral function $A(\omega,\mathbf{Q}+\mathbf{k}) =  \Im D^R(\omega,\mathbf{Q}+\mathbf{k})/\pi$ are substituted by a continuum of over-damped modes:
\begin{equation}\label{eq:PhotonSpectralDensity}
    A(\omega,\mathbf{Q}+\mathbf{k}) \sim \frac{[\omega+|E(\mathbf{k})|]^\alpha-[|E(\mathbf{k})|-\omega]^\alpha}{[(\omega+|E(\mathbf{k})|)^\alpha-(|E(\mathbf{k})|-\omega)^\alpha]^2 + b^2 E(\mathbf{k})^2},
\end{equation}
where $\alpha = 3/2$ for the square lattice and $\alpha = 1/2$ for the honeycomb. The distinct power-law dependence in (\ref{eq:PhotonSpectralDensity}) at small $\omega$ can be an experimental signature of the critical electromagnetic fluctuations affecting the electrons.
While such frequency- and momentum-resolved cavity probes are available in state-of-the-art experiments with atomic gases at the relevant frequencies and momenta \cite{mottl2012roton}, they seem rather challenging in the solid-state case, since the momentum scale $Q\sim p_F$ is much larger than the characteristic photon wave vectors.

\emph{Conclusions.---}
We have shown that cavity QED with 2D materials allows to implement and probe non-Fermi-liquid behavior under pristine conditions: 1) the underlying critical bosonic fluctuations which destroy the electronic quasiparticles are provided by the electromagnetic vacuum field and are thus controllable via cavity engineering; 2) 2D materials (or the synthetic versions based on ultracold gases) allow for enhanced tuneability of electronic properties and control over unwanted effects.

Our work introduces a new (experimentally relevant) model for the emergence of non-Fermi-liquid behavior, thus offering a new playground for controlled experimental investigations as well as for theoretical approaches.

Future studies shall provide improved theories for the superradiant criticality and the associated non-Fermi-liquid behavior (also including the unit-filling case on the square lattice where logarithmic divergences appear in the polarization), possibly identifying deviations from the universality class predicted for non-commensurate charge-density wave \cite{altshuler_1994,Holder2014} or Fulde-Ferrel-Larkin-Ovchinnikov \cite{piazza_2016,pimenov_2018} quantum critical points.

\emph{Acknowledgements.---}
We thank Ahana Chakraborty and Bernhard Frank for discussions.


%

\end{document}


\title{Supplemental Materials: Non-Fermi-liquid behavior from critical electromagnetic vacuum fluctuations}
	\author{Peng Rao}
	\affiliation{Max Planck Institute for the Physics of Complex Systems, N\"othnitzerstr. 38, 01187 Dresden, Germany}
	\author{Francesco Piazza}
	\affiliation{Max Planck Institute for the Physics of Complex Systems, N\"othnitzerstr. 38, 01187 Dresden, Germany}
	\date{\today}
	\begin{abstract}
		\tableofcontents
	\end{abstract}
	
	\maketitle
	
\section{Quantization of Electromagnetic Field in a cavity}\label{Appendix:CavityQuantisation}
	
	In this Appendix, we review the quantization of electromagnetic field in a cavity and derive the Green's function for cavity photons Eq.~(3) in the main text~\cite{Kakazu1994}. We begin by considering arbitrary cavity geometry. 
	
	We choose the Coulomb gauge for the four-potential operator $A_\mu = (\phi, \mathbf{A})$:
	\begin{equation}\label{eq:CoulombGauge}
		\text{div} \ \mathbf{A}=0.
	\end{equation}
	Under this gauge, the scalar potential $\phi$ mediates the direct Coulomb interaction between electrons. The electric and magnetic fields are then given by:
	\[
	\mathbf{E} = - \frac{\p \mathbf{A}}{\p t} - \nabla \phi, \ \mathbf{B} = \nabla \times \mathbf{A}.
	\] 
	
	The vector potential operator has the form:
	\begin{equation}\label{eq:VectorPotentialDef}
		\mathbf{A}= \sum_{a,n,k} \bigg(b^{a}_{n} \mathbf{A}^{a}_{n} e^{-i\omega_n t}+  b^{a\dagger}_{n} \mathbf{A}^{a*}_{n}e^{i \omega_n t}\bigg)
	\end{equation}
	Here $a$ and $n$ are indices for polarisation and photon energy levels $\omega_n$, and $b^a_{n}$ is the corresponding annihilation operator for the photons. A similar expression exists for $\phi$. Under the gauge (\ref{eq:CoulombGauge}), the equation of motion for the eigenfunctions $\mathbf{A}^{a}_{n}$ is:
	\begin{equation}\label{eq:PhotonEquation}
		\omega^2_{n}\mathbf{A}^{a}_{n} +\Delta\mathbf{A}^{a}_{n}=0,
	\end{equation}
	where $\Delta$ is the Laplacian. The scalar potential $\phi$ satisfies $\Delta \phi = 0$.

	Eq.~(\ref{eq:PhotonEquation}) must be supplemented by certain boundary conditions on the surface of the cavity. We shall assume that the cavity mirrors are perfectly conducting and there is no energy loss from the cavity: the surface impedance $\zeta=\surd(\mu/\varepsilon)=0$~\footnote{
		In principle, $\zeta(\omega)$ is a function of frequency. Here the approximation is that $\zeta(\omega)\approx 0$ in the frequency range of interest. 
	}.
	The boundary conditions are that the tangent components of $\mathbf{E}$ and the normal component of $\mathbf{H}$ to the conducting surface vanish there~\cite{Landau8}:
	\begin{equation}\label{eq:BoundaryConditions}
		E_{\parallel}= 0,~B_{\perp}=0.
	\end{equation}
	However, Eq.~(\ref{eq:BoundaryConditions}) is formuated in physical gauge-invariant quantities $\mathbf{E}$ and $\mathbf{B}$. The boundary conditions in terms of $\phi$, $\mathbf{A}$ are not unique. We shall choose the following boundary conditions which separate $\phi$ that mediates direct Coulomb interactions between electrons, and the vacuum cavity mode given by $\mathbf{A}$:
	\begin{equation}\label{eq:BoundaryConditions1}
		A_\parallel = 0,\ \phi = \text{const.}, \ (\nabla \times \mathbf{A})_{\perp}=0.
	\end{equation}
	This is because in the absence of matter, one can impose the further condition $\phi = 0$. Then the boundary condition~(\ref{eq:BoundaryConditions}) for the vacuum cavity modes becomes (\ref{eq:BoundaryConditions1}). In this paper we shall neglect effects from $\phi$ and are mainly concerned with interactions mediated by $\mathbf{A}$.

	Additionally, $\mathbf{A}^{a}_{n}$ satisfies the following normalization condition:
	\begin{equation}\label{eq:Normalisation}
		\int\mathbf{A}^{a*}_{n}.\mathbf{A}^{b}_{m} dV=\frac{1}{2\omega_{n}}\delta_{nm}\delta_{ab}.
	\end{equation} 
	Eq.~(\ref{eq:Normalisation}) is necessary for correctly quantizing the photon Hamiltonian:
	\begin{equation}\label{eq:PhotonHamiltonian}
		\begin{split}
			H = \int \frac{1}{2} \left( E^2+B^2\right) dV 
		\end{split}
	\end{equation}
	where the volume integral is taken over the entire cavity. Using the gauge condition (\ref{eq:CoulombGauge}) and the vector identity: 
	\[
	\begin{split}
		&(\nabla \times \mathbf{A})^2 = (\nabla \times \mathbf{A}).\mathbf{H} = \nabla.(\mathbf{A}\times\mathbf{H})+\mathbf{A}.(\nabla \times \mathbf{H});\\
		&\mathbf{A}.(\nabla \times \mathbf{H})= \mathbf{A}.[\nabla \times (\nabla \times \mathbf{A})]=  - \mathbf{A}.\Delta\mathbf{A},
	\end{split}
	\]
	Eq.~(\ref{eq:PhotonHamiltonian}) is transformed as:
	\begin{equation}\label{eq:PhotonHamiltonian1}
		\begin{split}
			H= \int \frac{1}{2} \bigg[ \bigg(\frac{\p \mathbf{A}}{\p t}\bigg)^2 + (\nabla \phi)^2- \mathbf{A}.\Delta\mathbf{A}\bigg]dV+ \int \left[\mathbf{A}\times (\nabla \times \mathbf{A})+ \phi \frac{\p \mathbf{A}}{\p t}\right].d\mathbf{S}.
		\end{split}
	\end{equation}
	The second integral is taken over the cavity surface and corresponds to energy loss to the cavity. With the perfect conductor boundary condition (\ref{eq:BoundaryConditions1}), this term vanishes as it should. Substituting Eqs.~(\ref{eq:VectorPotentialDef}) and (\ref{eq:PhotonEquation}), and the normalization condition (\ref{eq:Normalisation}) into (\ref{eq:PhotonHamiltonian1}) then gives:
	\begin{equation}\label{eq:PhotonHamiltonian2}
		\begin{split}
			H&=\sum_{n,a} \bigg(b^{a\dagger}_{n}b^{a}_{n}+\frac{1}{2}\bigg) \omega_{n}.
		\end{split}
	\end{equation}
	Thus the normalization condition (\ref{eq:Normalisation}) gives the quantized photon Hamiltonian. Note that with finite energy loss, the surface integral in (\ref{eq:PhotonHamiltonian1}) introduces an imaginary, dissipative term to (\ref{eq:PhotonHamiltonian2}), as coherent cavity modes become ill-defined.

	Eqs.~(\ref{eq:PhotonEquation}), (\ref{eq:BoundaryConditions1}) and (\ref{eq:Normalisation}) allow one to find the vacuum photon modes for any cavity geometry. In this paper, the cavity is taken to be two perfectly conducting planes perpendicular to the $z$-axis at $z=0, L$. Let us now find the corresponding eigenfunctions $\mathbf{A}^{a}_{n}$. Since the $(x,y)$-plane is infinite, $\mathbf{A}^{a}_{n}$ depends on the two-dimensional position vector $\mathbf{r} = (x,y)$ via $\exp(i \mathbf{k}.\mathbf{r})$.
	Denoting from now on $n$ this discrete index, we seek the solution in the form:
	\begin{equation}\label{eq:VectorPotential}
		\mathbf{A}(t,\mathbf{r},z)=\sum_{a,n,k} \frac{1}{\sqrt{2\omega_{nk}L} }\bigg(\mathbf{e}^{(a)}_{n} b^{a}_{nk} e^{-ikx} +  \mathbf{e}^{(a)*}_n b^{a\dagger}_{nk} e^{ikx} \bigg).
	\end{equation}
	Here $n$ is the index due to the finite boundary condition along the $z$-axis, $k x = \omega t - \mathbf{k}.\mathbf{r}$. The polarization vector $\mathbf{e}^{(a)}_n$ depends in general on $z$, $\mathbf{k}$ and $n$. 
	
	We take the direction of $\mathbf{k}$ to be along the $x$-axis so that $\mathbf{A}$ does not depend on $y$. To find the two physical polarizations, we first substitute the gauge condition (\ref{eq:CoulombGauge}), which gives:
	\begin{equation}\label{eq:GaugeCondition}
		\mathbf{k}.\mathbf{e}^{(a)} -i \frac{\p e^{(a)}_z}{\p z}=0.
	\end{equation}
	The first solution corresponds to the transverse cavity mode: $\mathbf{e}^{(1)}$ is along the $y$-axis. The boundary condition (\ref{eq:BoundaryConditions1}) and Eq.~(\ref{eq:Normalisation}) then gives:
	\begin{equation}\label{eq:PhotonPolarisation}
		\mathbf{e}^{(1)}_n = \sqrt{2}\mathbf{e}_y\sin\bigg(\frac{n\pi z}{L}\bigg),~n=1, 2...
	\end{equation}
	For the magnetic field:
	\[
	B_y=0, B_z  =\frac{\p A_y}{\p x},~B_x = -\frac{\p A_y}{\p z}.
	\]
	Then $B_z=0$ is also satisfied at the boundary. Substituting our solution into Eq.~(\ref{eq:PhotonEquation}) gives for the cavity photon dispersion:
	\begin{equation}\label{eq:PhotonDispersion}
		\omega_{nk}^2 = k^2 + \omega_{0n}^2,\ \omega_{0n} =\frac{n\pi z}{L}.
	\end{equation}
	Since $n \ne 0$, the photon dispersion is always gapped.
	
	The second polarisation lies in the $(x,z)$-plane and $A_y=0$. For completeness we give its solution, although the mode is not used in the main text. We have for the magnetic field:
	\[
	B_x = 0 ,~B_y= \frac{\p A_x}{\p z} - \frac{\p A_z}{\p x},~B_z = 0.
	\]
	Then $B_z=0$ automatically. The boundary conditions for $\mathbf{A}$ in (\ref{eq:BoundaryConditions1}) gives:
	\[
	A_x\propto \sin\bigg(\frac{n\pi z}{L}\bigg).
	\]
	We then have from the gauge condition (\ref{eq:GaugeCondition}):
	\[
	kA_x = i \frac{\p A_z}{\p z} \rightarrow A_z \propto \cos \bigg(\frac{n\pi z}{L}\bigg),
	\]
	and the cavity photon dispersion is also given by (\ref{eq:PhotonDispersion}). The polarisation vector satisfies:
	\[
	\begin{split}
		&\mathbf{e}^{(2)}_n = a\sin\bigg(\frac{n\pi z}{L}\bigg) \mathbf{e}_x+b\cos \bigg(\frac{n\pi z}{L}\bigg)\mathbf{e}_z ,\\
		&|a|^2+|b|^2=2,~b=\frac{ikL}{n\pi} a.
	\end{split}
	\]
	The second and third equalities are obtained by substituting (\ref{eq:Normalisation}) and (\ref{eq:CoulombGauge}). This gives (without loss of generality $a$ can be taken real):
	\[
	a = \bigg[\frac{2}{1+ (kL/n\pi)^2}\bigg]^{1/2}.
	\]
	The electrons are placed at $z=L/2$. Note that in this case $n=0$ can be taken and the photon mode is gapless. But then $A_x=A_y=0$ and the system does not couple to two-dimensional currents. 
	
	We are now in a position to calculate the photon's Green's function. Since the electrons are placed at the center between two mirrors, we shall always take $z=L/2$ for $\mathbf{A}$ in what follows. Denoting $\mathbf{r}$ to be the two-dimensional in-plane position vector, the free photon Green's function is:
	\begin{equation}
		D_{0,ik}(t,\mathbf{r}) =i \bra{0} T A_i\left(t,\mathbf{r},L/2\right)A_k\left(0,0,L/2\right) \ket{0},
	\end{equation}
	where $T$ is time-ordering. For our purpose in the main text, we retain only the contribution from the transverse modes at $n=1$. Substituting then (\ref{eq:VectorPotential}) and using the identities:
	\[
	\bra{0}  b^{a}_{k} b^{b\dagger}_{k'} \ket{0} =\delta_{ab} \delta(\mathbf{k}-\mathbf{k}'),
	\]
	we obtain:
	\[
	\begin{split}
		&D_{0,ik}(t,\mathbf{r}) = \frac{2}{L} \sum_{\mathbf{k}}e_i(\mathbf{k}) e_k^*(\mathbf{k}) D_0(t,\mathbf{k}), \ \mathbf{e}(\mathbf{k}).\mathbf{k}=0, \\
		&D_0(t,\mathbf{k}) =	 \frac{i}{2\omega_k}
		\begin{cases} 
			\exp(-i\omega_k t),~t>0;\\
			\exp(i\omega_k t),~t<0,
		\end{cases}
	\end{split}
	\]
	where $\mathbf{e}(\mathbf{k})$ is the unit transverse polarization vector and $\omega_k = k^2 + \omega_0^2$, $\omega_0 = \pi /L$. Taking the Fourier transform of $D(t,\mathbf{k})$ and regularizing the integral over $t$ by displacing $\omega_k \rightarrow \omega_k \mp i0$ at $t \rightarrow \pm \infty$ then gives:
	\begin{equation}\label{eq:PhotonPropagator}
		D_{0,ik}(k) = -\frac{2}{L} e_i(\mathbf{k}) e_k^*(\mathbf{k}) \frac{1}{\omega^2-\omega_k^2+ i0}.
	\end{equation}
	This is Eq.~(3) in the main text. Note that contrary to the longitudinal polarization, the transverse mode is always in-plane for arbitrary $\mathbf{k}$ and does not depend on $n$.

	\section{The case with multiple cavity modes}
	
	Let us discuss the modifications to the results in this paper, when multiple cavity modes are taken into account. This is necessary for solid state systems, in which the characteristic momentum of photons necessary for non-FL scaling exceeds the cavity photon gaps $\omega_{0n}$. In this case the superradiance condition below Eq.~(5) in the main text is modified, but this does not affect the non-FL scaling. This is because the overdamped polaritonic modes at the superradiance critical point are essentially electronic fluctuations, which are in a sense independent from the free photon dispersions.

	To be more precise, we introduce explicitly the $z$-coordinate dependence in the photon Hamiltonian and keep electron fields $\Psi$ purely two-dimensional:
	\[
	H = \int \Psi^\dagger(\mathbf{r}) \varepsilon(\hat{p}) \Psi(\mathbf{r}) d^2x + \frac{1}{2}\int (E^2+B^2) d^2x dz + V,
	\]
	with electron-photon coupling:
	\[
	V = - \int \mathbf{j}(\mathbf{r}).\mathbf{A}(\mathbf{r},z) \delta(z-L/2) d^2x dz.
	\]
	The current and electron fields depend only on $\mathbf{r}=(x,y)$. From Eq.~(\ref{eq:PhotonPolarisation}), the real eigenfunctions for the transverse photon modes along the $z$-axis are:
	\begin{equation}\label{Eq:PhotonEigenfunctions}
		\phi^{(0)}_n(z) = \sqrt{\frac{2}{L}} \sin \left(\frac{n\pi z}{L}\right).
	\end{equation}

	The Dyson's equation for the transverse photon propagator with in plane momenta $\mathbf{k}$ and $z$-coordinates (all quantities are the retarded functions):
	\begin{equation}\label{Eq:DysonsEquationMultiModes}
		D(\omega,\mathbf{k};z,z') = D_0(\omega,\mathbf{k};z,z') + \int D_0(\omega,\mathbf{k};z,z_1)\Pi(\omega,\mathbf{k};z_1,z_2)D(\omega,\mathbf{k};z_2,z') dz_1 dz_2.
	\end{equation}
	The photon polarization function has the form:
	\[
	\Pi(\omega,\mathbf{k};z_1,z_2) = \Pi(\omega,\mathbf{k}) \delta(z_1-L/2)\delta(z_2-L/2),
	\]
	where $\Pi(\omega,\mathbf{k})$ is the two-dimensional transverse polarization. In terms of photon eigenmodes, $D, D_{0}$ and $\Pi$ become matrices in cavity mode indices. To obtain the Dyson equation between them, we multiply Eq.~(\ref{Eq:DysonsEquationMultiModes}) by $\phi^{(0)}_{n}(z) \phi^{(0)}_m(z')$ and integrating over $z$, $z'$: 
	\[
	\hat{D}(\omega,\mathbf{k}) = \hat{D}_{0}(\omega,\mathbf{k}) +  \hat{D}_{0}(\omega,\mathbf{k})\hat{\Pi}(\omega,\mathbf{k})\hat{D}(\omega,\mathbf{k}),
	\]
	where $\hat{D}_0$ and $\hat{\Pi}$ are given by:
	\begin{align}
		[\hat{D}_{0}]_{nm}(\omega,\mathbf{k}) &= D^{(n)}_{0}(\omega,\mathbf{k})  \delta_{nm} , \  D^{(n)}_{0}(\omega,\mathbf{k}) = -\frac{2}{L} \frac{1}{\omega^2 - \omega_{nk}^2}; \\
		\hat{\Pi}_{nm}(\omega,\mathbf{k}) & =\Pi(\omega,\mathbf{k}) \phi^{(0)}_{n}(L/2) \phi^{(0)}_m(L/2).
	\end{align}
	Note that even modes decouple from electronic fluctuations at $z=L/2$ since $\phi_{2n}(L/2)$ vanishes identically. For odd modes, $\phi^{(0)}_{2n+1}(L/2)= \sqrt{2/L} (-1)^{n}$. Thus apart from an overall sign, the polarisation matrix is the same for all odd indices (here it is important that the transverse polarization vector does not depend on $n$). Retaining only odd indices in what follows, we obtain:
	\begin{equation}\label{Eq:DysonsEquationMultiModes1}
		[\hat{D}(\omega,\mathbf{k})]^{-1} = [\hat{D}_{0}(\omega,\mathbf{k})]^{-1} - \hat{\Pi}(\omega,\mathbf{k}).
	\end{equation}
	Eq.~(\ref{Eq:DysonsEquationMultiModes1}) gives the photon propagator matrix in the case of multiple modes.

	The superradiance critical point condition can now be formulated as the critical coupling at which one of the eigenvalues of $[\hat{D}(0,\mathbf{Q})]^{-1}$ first becomes zero; see Ref.~\cite{Andolina2020} which also discusses the feasibility of superradiance in solid state systems. Denote the corresponding eigenvector of $\hat{D}$ at the superradiance critical point as $v^{(\alpha)}$, the condition becomes:
	\[
	[D^{(\alpha)}(0,\mathbf{Q})]^{-1} = \sum_{m,n} [D_{0}(0,\mathbf{Q})]^{-1}_{nm} v^{(\alpha)}_n v^{(\alpha)}_m - C_\alpha\Pi(0,\mathbf{Q})=0; \  C_\alpha = \left( \sum_{n} \phi^{(0)}_{n}(L/2) v^{(\alpha)}_{n} \right)^2.
	\]
	Here $n,m$ are odd cavity indices. It can then be seen that the expansion of $D^{(\alpha)}(\omega,\mathbf{Q}+\mathbf{k})$ at small $\omega$ and $\mathbf{k}$ will contain the same non-analytical terms from $\Pi(\omega,\mathbf{Q}+\mathbf{k})$ as in the main text.
	
	The electron self-energy is given by:
	\begin{equation}\label{Eq:SelfEnergyMultiModes}
		\begin{aligned}
			\Sigma\left(p\right)& =T \sum_{k_0}  \int \delta(z-L/2) \delta(z'-L/2)  G(p+k) D(k;z,z') dz dz'\frac{d^2k}{(2\pi)^3}\\
			&= \ T \sum_{k_0} \int \sum_l C_l  D^{(l)}(k) G\left(p+k\right)\frac{d^2k}{(2\pi)^3}.
		\end{aligned}
	\end{equation}
	Here $k_0$ is the bosonic Matsubara frequency and $l$ denotes the eigenmode indices of $\hat{D}$. Thus each eigenmode scatters independently with electrons. To consider the non-FL effect one only needs to retain in (\ref{Eq:SelfEnergyMultiModes}) the superradiant mode $\alpha$. It can see that, apart from an overall constant, the calculations are identical to the ones given below in this Supplemental Material for the one mode case, and non-FL scaling is unaffected.

	\section{Computation of Photon Polarization function on the square lattice}
	
	In this section we compute Eq.~(9) in the main text. The polarization function is given by the diagram in Fig.~2(a) in the main text:
	\begin{equation}\label{eq:PolarizationSquare}
		\Pi(\omega_n,\mathbf{Q}+\mathbf{k}) = 2a^4(et)^2 T\sum_m \int  \bigg(p_\parallel+\frac{k_\parallel}{2}\bigg)^2 G(\nu_m,\mathbf{p}) G(\omega_n+ \nu_m, \mathbf{p}+\mathbf{Q}+\mathbf{k})\frac{d^2p}{(2\pi)^2}.
	\end{equation}
	In Eq.~(\ref{eq:PolarizationSquare}) the sum over Matsubara frequency is first carried out. This gives:
	\begin{equation}\label{eq:PolarizationSquare-1}
		- \frac{2a^4(et)^2}{(2\pi)^2} \int \left[p_\parallel+\frac{k_\parallel}{2}\right]^2\int\bigg(n_{+}(p+k)-n_{-}(p)\bigg) \frac{d^2p}{i\omega_n - [\xi_{+}(p+k)-\xi_{-}(p)]}d^2p,
	\end{equation}
	where $\xi_{\pm}(k)= k_\parallel^2/(2m) \pm v_Fk_\perp$ is the dispersion near the `hot-spots' and $n_{\pm}(p)$ is the Fermi distribution for energy $\xi_{\pm}(p)$. For the first term in the bracket, we perform the substitution $p+k\rightarrow p$ and take $T\rightarrow0$:
	\begin{equation}
		\begin{split}
			&- \frac{2a^4(et)^2}{(2\pi)^2} \int dp_\parallel \int_{-\Lambda_0}^{-p_\parallel^2/(2mv_F)} \bigg(p_\parallel-\frac{k_\parallel}{2}\bigg)^2 \frac{dp_\perp}{i\omega_n+v_Fk_\perp+k_\parallel^2/(2m) - (2v_Fp_\perp+p_\parallel k_\parallel/m)} \\
			\propto & \int_{-\Lambda_0}^{\Lambda_0}\bigg(p_\parallel-\frac{k_\parallel}{2}\bigg)^2  \log\left[i\omega_n+v_Fk_\perp+\frac{k_\parallel^2}{4m}+ \frac{(p_\parallel-k_\parallel/2)^2}{m}\right]dp_\parallel.
		\end{split}
	\end{equation}
	The integral over $p_\parallel$ is formally divergent. Thus we need to extract from the following integral the contribution close to the Fermi-surface:
	\[
	\begin{split}
		\int x^2 \log(x^2+a) dx=\frac{x^3}{3}\log(x^2+a) - \frac{2}{9} x^3 + \frac{2}{3} a x - \frac{2}{3}\int \frac{a^2}{x^2+a} dx; \ a = i\omega_n+v_Fk_\perp+\frac{k_\parallel^2}{4m}.
	\end{split}
	\]
	Only the last term converges near the Fermi surface to give $-2\pi a^{3/2}/3$. This gives:
	\[
	-\frac{(aet)^2}{6\pi v_F}m^{3/2} \bigg(v_Fk_\perp+\frac{k_\parallel^2}{4m}+i\omega_n\bigg)^{3/2}
	\]
	
	For the second term in \Eqref{eq:PolarizationSquare-1} we get:
	\begin{equation}
		\begin{split}
			&- \frac{(aet)^2}{(2\pi)^2} \int dp_\parallel \int^{\Lambda_0}_{p_\parallel^2/(2mv_F)} \bigg(p_\parallel+\frac{k_\parallel}{2}\bigg)^2 \frac{dp_\perp}{i\omega_n-v_Fk_\perp-k_\parallel^2/(2m) - (2v_Fp_\perp+p_\parallel k_\parallel/m)} \\
			=&\frac{1}{2(2\pi)^2v_F} \int_{-\Lambda_0}^{\Lambda_0}\bigg(p_\parallel+\frac{k_\parallel}{2}\bigg)^2  \log\bigg(v_Fk_\perp+\frac{k_\parallel^2}{4m}-i\omega_n+ \frac{(p_\parallel+k_\parallel/2)^2}{m}\bigg)dp_\parallel.
		\end{split}
	\end{equation}
	And we get:
	\[
	-\frac{(aet)^2}{6\pi v_F}m^{3/2} \bigg(v_Fk_\perp+\frac{k_\parallel^2}{4m}-i\omega_n\bigg)^{3/2}.
	\]
	Summing the two parts and performing analytical continuation $i\omega_n\rightarrow \omega+i0$ we return to Eq.~(9) in the main text.

	\section{Computation of Electron self-energy on the square lattice}
	
	Eq.~(12) in the main text [given by Fig.~2(c)] is computed as follows. Under the change of variables to $k', \omega'$:
	\[
	-k' = E(k),~ \omega' = \varepsilon(k) = \frac{k_\parallel^2}{2m}+ k_\perp.~J = 2m/k_\parallel = \sqrt{m/(\omega'+k')},~k_\parallel^2 = 4m(\omega'+k'),
	\]
	where $J$ is the Jacobian of the integral, and substituting $D^R$ from Eq.~(10), Eq.~(11) becomes:
	\begin{equation}\label{eq:SelfEnergy-1}
		\begin{split}
			\Im \Sigma \left(\omega,\frac{\mathbf{Q}}{2}\right) \propto & \int_0^\omega d\omega'\int_0^{\Lambda_0} dk'\frac{\sqrt{\omega'+k'}}{i (k'- \omega'+\omega )^{3/2}-i(k'+\omega'-\omega)^{3/2}+b k'+ c k'^2}.
		\end{split}
	\end{equation}
	The dominant contribution comes from $k'\gg \omega-\omega'>0$, at which the non-analytical term in the photon Green's function Eq.~(10) gives a large imaginary part that is responsible for Landau damping~\footnote{
		Here the convention for circumventing the branch-cuts in complex plane of $k'$ is: for $(\omega'-\omega -k'+i0)^{3/2}$ as $k'$ increases it goes from under the branch-cut to the real axis, whereas for $(\omega-\omega' -k'-i0)^{3/2}$ it goes from above. 
	}:
	\[
	\begin{split}
		&[-k' \pm (\omega'-\omega +i0)]^{3/2} \rightarrow  \pm i [k' \mp (\omega'-\omega +i0)]^{3/2};\\
		& i (k'- \omega'+\omega )^{3/2}-i(k'+\omega'-\omega)^{3/2}\approx 3i(\omega-\omega')\sqrt{k'}.
	\end{split}
	\]
	This gives for Eq.~(\ref{eq:SelfEnergy-1}):
	\begin{equation}
		\begin{split}
			\Im \Sigma \left(\omega,\frac{\mathbf{Q}}{2}\right) \propto \int_0^\omega d\omega'\int_0^{\Lambda_0} dk'\frac{(\omega-\omega')}{(\omega-\omega')^2+(b+ck')^2k'}.
		\end{split}
	\end{equation}
	Here we have defined new constants $b,c\rightarrow b/3,c/3$. The upper limit is at $\Lambda_0$, since the expressions for $D^R$ in Eq.~(10) and self-energy (11) in the main text no longer hold at large $k'\sim  \Lambda_0$. Another change into dimensionless variables:
	\[
	\omega'=\omega \omega_1,~k'=\omega k_1,
	\]
	gives for the self-energy:
	\begin{equation}\label{eq:SelfEnergy-2}
		\omega \int_0^1 d\omega_1\int_0^{\sim\Lambda_0 /\omega} dk'\frac{(1-\omega_1)}{(1-\omega_1)^2+(b/\sqrt{\omega}+ck_1\sqrt{\omega})^2k_1}.
	\end{equation}
	The lower limit for $k'$ is strictly speaking not zero but a term of the magnitude $\omega-\omega'$. This substitution then results in an error of $\omega^2$ which is not important here. In the second bracket in the denominator of (\ref{eq:SelfEnergy-2}), the first term is dominant at $k_1\ll b/(\omega c)$. But since $b$ and $c$ comes from integration far from the Fermi-surface, the only momentum scale is $\Lambda_0$. Hence by dimensional arguments $b/(\omega c)\sim \Lambda_0/\omega$,
	which is just the upper-limit in (\ref{eq:SelfEnergy-2}). Therefore we can set $c=0$ in (\ref{eq:SelfEnergy-2}) and obtain, up to logarithmic accuracy:
	\begin{equation}
		\Im \Sigma\left( \omega, \frac{\mathbf{Q}}{2}\right) \sim \omega^2\log |\omega|.
	\end{equation}
	This is Eq.~(12) in the main text.
	
	Note that the self-energy for the electrons on a honeycomb lattice in Eq.~(17) corresponds to formally setting $b=0$ in Eq.~(\ref{eq:SelfEnergy-2}). The integral is then convergent and gives:
	\begin{equation}
		\begin{split}
			\Im \Sigma \left(\omega,\frac{\mathbf{Q}}{2}\right)  \sim \omega \int_{0}^{1} d\omega_1 \int_{0}^{\infty} \frac{1-\omega_1}{(1-\omega_1)^2 + c^2\omega {k}^3_1}d\omega_1dk_1\\
			= \bigg(\frac{\omega}{c}\bigg)^{2/3}  \int_{0}^{1} (1-\omega_1)^{-1/3}d\omega_1 \int_{0}^{\infty} \frac{dx}{1+x^3} = \frac{\pi}{\sqrt{3}}\bigg(\frac{\omega}{c}\bigg)^{2/3}.
		\end{split}
	\end{equation}
	Here the dominant contribution comes from $k'\sim \Lambda_0 (\omega/\Lambda_0)^{2/3}\gg \omega$.

	\section{Computation of Polarization function on the honeycomb lattice}
	
	On the honeycomb lattice, the free electron Green's function for the Hamiltonian in Eq.~(13) is a matrix in spin and pseudo-spin space:
	\begin{equation}\label{eq:GrapheneGreensFunction}
		\begin{split}
			&G_{\alpha\beta}(\nu_n,\mathbf{k}) =G(\nu_n,\mathbf{k})  \delta_{\alpha\beta},\\
			&G(\omega_n,\mathbf{k})= \sum_{\lambda} G_\lambda(\omega_n,\mathbf{k}) u_\lambda(\mathbf{k})u_\lambda^*(\mathbf{k}), \\
			&G_\lambda(\omega_n,\mathbf{k}) = \frac{1}{ i\omega_n - \varepsilon_{\lambda}(k) + \mu}.
		\end{split}
	\end{equation}
	$G_\lambda(\nu_n,\mathbf{k})$ are the Green's functions of the electron and hole bands for $\lambda= \pm$ respectively. We rewrite Eq.~(\ref{eq:GrapheneGreensFunction}) by substituting Eq.~(14) in the main text to give:
	\begin{equation}\label{eq:GrapheneGreensFunction1}
		\begin{split}
			G(\nu_n,\mathbf{k}) = \frac{1}{2}\bigg(\left[G_+(\nu_n,\mathbf{k})+G_-(\nu_n,\mathbf{k})\right] +\bm{\sigma}.\mathbf{n}\left[G_+(\nu_n,\mathbf{k})-G_-(\nu_n,\mathbf{k})\right]\bigg).
		\end{split}
	\end{equation}
	Here $\mathbf{n}$ is the directional vector of $\mathbf{k}$.
	
	To leading order, the polarization tensor $\Pi_{\mu\nu} (\omega_n,\mathbf{Q}+\mathbf{k})$ in Fig.~2(a) in the main text near $\mathbf{Q}$ is proportional to:
	\begin{equation}\label{eq:GraphenePolarisation}
		\begin{split}
			T\sum_m \int  \Tr \left[ \sigma_\mu G(\nu_m,\mathbf{p})
			\sigma_\nu G(\omega_n+ \nu_m, \mathbf{p}+\mathbf{Q}+\mathbf{k})  \right]\frac{d^2p}{(2\pi)^2}.
		\end{split}
	\end{equation}
	To obtain Eq.~(16) in the main text, we substitute Eq.~(\ref{eq:GrapheneGreensFunction1}) into (\ref{eq:GraphenePolarisation}) and project to the transverse photon mode. The result is simplified using the Pauli matrix identities (terms with the anti-symmetric tensor $e_{\mu\nu \delta}$ vanish since $\mu, \nu = 1, 2$):
	\[
	\begin{split}
		\Tr (\sigma_a\sigma_b) &=2 \delta_{ab}, \ \Tr (\sigma_a\sigma_b\sigma_c\sigma_d) = 2(\delta_{ab}\delta_{cd}-\delta_{ac}\delta_{bd}+\delta_{ad}\delta_{bc}),
	\end{split}
	\]
	then contracted with:
	\[
	e_\mu(\mathbf{Q}+\mathbf{k})e_\nu(\mathbf{Q}+\mathbf{k}) \approx e_\mu(\mathbf{Q})e_\nu(\mathbf{Q}).
	\]
	After some algebra we obtain for the $G_+G_+$ term in $\Pi (\omega_n,\mathbf{Q}+\mathbf{k})$ up to terms of order $k^2/p_F^2$,:
	\begin{equation}
		2e^2v_F^2T\sum_m\int G_+(\nu_m+\omega_n,\mathbf{p}+\mathbf{Q}+\mathbf{k})G_+(\nu_m,\mathbf{p}) \frac{d^2p}{(2\pi)^2}
	\end{equation}
	The integral has the form:
	\begin{equation}\label{eq:GraphenePolarisation-1}
		\begin{split}
			&\sum_m\int G_+(\nu_m+\omega_n,\mathbf{p}+\mathbf{Q}+\mathbf{k})G_+(\nu_m,\mathbf{p}) \frac{d^2p}{(2\pi)^2} \\
			=&\sum_m\int  \frac{d^2p}{(2\pi)^2}  \left(i\nu_m+ v_Fp_\perp -\frac{p_\parallel^2}{2m} \right)^{-1}
			\left[i(\nu_m+\omega_n) - v_F(p_\perp+k_\perp)-\frac{(p_\parallel+k_\parallel)^2}{2m}\right]^{-1}.
		\end{split}
	\end{equation}
	This is calculated analogously to Eq.~(\ref{eq:PolarizationSquare}) by summing over Matsubara frequency $\nu_m$ then taking $T\rightarrow 0$. This gives:
	\begin{equation}\label{eq:GraphenePolarisation-2}
		\begin{split}
			&-\frac{1}{(2\pi)^2} \int\bigg(n_{+}(p+k)-n_{-}(p)\bigg) \frac{d^2p}{i\omega_n - [\xi_{+}(p+k)-\xi_{-}(p)]}\\
			&=- \frac{1}{(2\pi)^2} \int \bigg(\frac{n_{+}(p)}{i\omega_n - [\xi_{+}(p)-\xi_{-}(p-k)]}- \frac{n_{-}(p)}{i\omega_n - [\xi_{+}(p+k)-\xi_{-}(p)]}\bigg)d^2p.
		\end{split}
	\end{equation}
	At $T=0$, we consider the first term in the bracket in (\ref{eq:GraphenePolarisation-2}) and integrate over $p_\perp$:
	\begin{equation}\label{eq:GraphenePolarisation-3}
		\begin{split}
			&- \frac{1}{(2\pi)^2} \int dp_\parallel \int_{-\Lambda_0}^{-p_\parallel^2/(2mv_F)} \frac{dp_\perp}{i\omega_n+v_Fk_\perp+k_\parallel^2/(2m) - (2v_Fp_\perp+p_\parallel k_\parallel/m)} \\
			=&\frac{1}{2(2\pi)^2v_F} \int_{-\Lambda_0}^{\Lambda_0}  \log\left(i\omega_n+v_Fk_\perp+\frac{k_\parallel^2}{4m}+ \frac{(p_\parallel-k_\parallel/2)^2}{m}\right)dp_\parallel.
		\end{split}
	\end{equation}
	Here $\Lambda_0$ is a cut-off of the order of the bandwidth. In calculating the integral over the logarithm we extract the contribution from close to the Fermi-surface in the same way as in the square lattice case:
	\[
	\int \log(x^2+a) dx =2\Lambda_0 \log (\Lambda_0^2+a) -4\Lambda_0 +2a \int \frac{1}{x^2+a} dx;\ a = i\omega_n+v_Fk_\perp+\frac{k_\parallel^2}{4m}.
	\]
	The logarithmic terms can be expanded in powers of $a$ which would give far from Fermi surface corrections. The last term is convergent and we have:
	\begin{equation}
		\bigg(v_F k_\perp +\frac{k_\parallel^2}{4m}+i\omega_n \bigg)^{1/2}.
	\end{equation}
	The second term in (\ref{eq:GraphenePolarisation-2}) gives:
	\begin{equation}
		\bigg(v_F k_\perp +\frac{k_\parallel^2}{4m}-i\omega_n \bigg)^{1/2}.
	\end{equation}
	Adding them together and performing the analytical continuation $i\omega_n \rightarrow \omega + i0$ gives Eq.~(16) in the main text.

	
%